# A meshless, integration-free, and boundary-only RBF technique

W. Chen[*] and M. Tanaka

Department of Mechanical System Engineering, Shinshu University, Wakasato 4-17-1, Nagano City, Nagano, JAPAN (Present email:[*] mawchen@cityu.edu.hk )

## Abstract

Based on the radial basis function (RBF), non-singular general solution and dual reciprocity method (DRM), this paper presents an inherently meshless, integration-free, boundary-only RBF collocation techniques for numerical solution of various partial differential equation systems. The basic ideas behind this methodology are very mathematically simple. In this study, the RBFs are employed to approximate the inhomogeneous terms via the DRM, while non-singular general solution leads to a boundary-only RBF formulation for homogenous solution. The present scheme is named as the boundary knot method (BKM) to differentiate it from the other numerical techniques. In particular, due to the use of nonsingular general solutions rather than singular fundamental solutions, the BKM is different from the method of fundamental solution in that the former does no require the artificial boundary and results in the symmetric system equations under certain conditions. The efficiency and utility of this new technique are validated through a number of typical numerical examples. Completeness concern of the BKM due to the only use of non-singular part of complete fundamental solution is also discussed.

# 1. Introduction

It has long been claimed that the boundary element method (BEM) is a viable alternative to the domain-type finite element method (FEM) and finite difference method (FDM) due to its advantages in dimensional reducibility and suitability to infinite domain problems. However, nowadays the FEM and FDM still dominate science and engineering computations. The major bottlenecks in performing the BEM analysis have long been its weakness in handling inhomogeneous terms such as time-dependent and nonlinear problems. The recent introduction of the dual reciprocity BEM (DRBEM) by Nardini and Brebbia [1] greatly eases these inefficiencies. Notwithstanding, as was pointed out in [2], the method is still more mathematically complicated and requires strenuous manual effort compared with the FEM and FDM. In particular, the handling of singular integration is not easy to non-expert users and often computationally expensive. The use of the low order approximation in the BEM also slows convergence. More importantly, just like the FEM, surface mesh or re-mesh in the BEM requires costly computation, especially for moving boundary and nonlinear problems. The method of fundamental solution (MFS) is shown an emerging technique to alleviate these drawbacks and is getting increasing attraction especially due to some recent works by Golberg, Chen and Muleskov [2-5]. The MFS affords advantages of integration-free, spectral convergence, and meshless. However, the use of artificial boundary outside physical domain has been a major limitation of the MFS, which may cause severe ill-conditioning of the resulting equations, especially for complex boundary geometry [6,7]. These inefficiencies of the MFS motivate us to find an alternative technique, which keeps its merits but removes its shortcomings undermining its attractiveness.

Recently, Golberg et al. [4] established the DRBEM on the firm mathematical theory of the radial basis function (RBF). The DRBEM can be regarded as a two-step methodology. In terms of dual reciprocity method (DRM), the RBF is applied at first to approximate the particular solution of inhomogeneous terms, and then the standard BEM is used to discretize the remaining homogeneous equation. Chen et al [3] and Golberg et al [2,4] extended this RBF approximation of particular solution to the MFS, which greatly enhances its applicability. In fact, the MFS itself can also be considered a special RBF collocation approach, where the fundamental solution of the governing equation is taken as the radial function. On the other hand, the domain-type RBF collocation is also now under intense study since Kansa's pioneer work [8]. The major charming of the RBF-type techniques is its meshless inherence. The construction of mesh in high dimension is not a trivial work. Unlike the recently developed meshless FEM with the moving least square, the RBF approach is truly cheap meshless technique without any difficulty applying boundary conditions [9, 10]. The RBF is therefore an essential component in this study to construct a viable numerical technique.

Kamiya et al. [11] and Chen et al. [12] pointed out that the multiple reciprocity BEM (MRM) solution of the Helmholtz problems with the Laplacian plus its high-order terms is in fact to employ only the singular real part of complete complex fundamental solution of the Helmholtz operator. Power [13] simply indicated that the use of either the real or imaginary part of the Helmholtz Green's representation formula could formulate interior Helmholtz problems. This study extends these ideas to general problems such as Laplace and convection-diffusion problems by a combined use of the nonsingular general solution, the dual reciprocity method, and RBF. This mixed technique is named as the boundary knot method (BKM) [14] due to its essential meshless property, namely, the BKM does not need any discretization grids for any dimension problems and only uses knot points. The inherent inefficiency of the MFS due to the use of the fictitious boundary is alleviated in the BKM, which leads to tremendous improvement in computational efficiency and produces the symmetric matrix structure under certain conditions. It is noted that the BKM dose not involve integration operation due to the use of the collocation technique. Just like the MFS, the method is very simple to implement. The nonsingular general solution for multidimensional problems can be understood the nonsingular part of complete fundamental solution of various operators. The preliminary numerical studies of this paper show that the BKM is a promising technique in terms of efficiency, accuracy, and simplicity. We also use the BKM with the response knot-dependent nonsingular general solutions to solve the varying parameter problems successfully.

This paper is organized as follows. Section 2 involves the procedure of the DRM and RBF approximation to particular solution. In section 3, we introduce nonsingular general solution and derive the analogization equations of the BKM. Numerical results are provided and discussed in section 4 to establish the validity and accuracy of the BKM. Completeness concern of the BKM is discussed in section 5. Finally, section 6 concludes some remarks based on the reported results. In appendix, we give the nonsingular general solution of some 2D, 3D steady and time-dependent operators.

## 2. RBF approximation to particular solution

Like the DRBEM and MFS, the BKM can be viewed as a two-step numerical scheme, namely, DRM and RBF approximation to particular solution and the evaluation of homogeneous solution. The latter is the emphasis of this paper. The former has been well developed now [2-4]. For the sake of completeness, we here outline the basic methodology to approximate particular solution. Let us consider the differential equation

$$L\{u(x)\} = f(x), \quad x \in \Omega \tag{1}$$

with boundary conditions

$$u(x) = b_1(x), \quad x \subset \Gamma_u, \tag{2}$$

$$\frac{\partial u(x)}{\partial n} = b_2(x), \quad x \subset \Gamma_T, \tag{3}$$

where $L$ is a differential operator, $f(x)$ is a known forcing function. $n$ is the unit outward normal. $x \in R^d$, $d$ is the dimension of geometry domain, which is bounded by a piece-wise smooth boundary $\Gamma = \Gamma_u + \Gamma_T$. In order to facilitate discussion, it is assumed here that operator $L$ includes Laplace operator, namely,

$$L\{u\} = \nabla^2 u + L_1\{u\}. \tag{4}$$

We should point out that this assumption is not necessary [15]. Eq. (1) can be restated as

$$\nabla^2 u + u = f(x) + u - L_1\{u\}. \tag{5}$$

The solution of the above equation (5) can be expressed as

$$u = v + u_p, \tag{6}$$

where $v$ and $u_p$ are the general and particular solutions, respectively. The latter satisfies equation

$$\nabla^2 u_p + u_p = f(x) + u - L_1\{u\}, \tag{7}$$

but does not necessarily satisfy boundary conditions (2) and (3). $v$ is the homogeneous solution of the Helmholtz equation

$$\nabla^2 v + v = 0, \quad x \in \Omega, \tag{8}$$

$$v(x) = b_1(x) - u_p, \quad x \subset \Gamma_u, \tag{9}$$

$$\frac{\partial v(x)}{\partial n} = b_2(x) - \frac{\partial u_p(x)}{\partial n}, \quad x \subset \Gamma_T. \tag{10}$$

The first step in the BKM is to evaluate the particular solution $u_p$ by the DRM and RBF. After this, Eq. (8)-(10) can be solved by the boundary RBF methodology using non-singular general solution proposed in later section 3.

Unless the right side of Eq. (7) is rather simple, it is practically impossible to get analytical particular solution in general cases. In addition, even if the analytical solutions for some problems are available, their forms are usually too complicated to use in practice. Therefore, we prefer to approximate these inhomogeneous terms numerically. The DRM with the RBF is a very promising approach for this task [1-5], which analogizes the particular solution by the use of a series of approximate particular solution at all specified nodes. The right side of Eq. (7) is approximated by the RBF approach, namely,

$$f(x) + u - L_1\{u\} \cong \sum_{j=1}^{N+L} \alpha_j \phi(\|x - x_j\|) + \psi(x), \tag{11}$$

where $\alpha_j$ are the unknown coefficients. $N$ and $L$ are respectively the numbers of knots on boundary and domain. $\|\ \|$ represents the Euclidean norm, $\phi(\ )$ is the RBF. Additional polynomial term $\psi(x)$ is required to assure nonsingularity of interpolation matrix if the RBF is conditionally positive definite such as multiquadratics (MQ) and thin plate spline (TPS) [8,16]. For example, in the 2D case with linear polynomial restrains we have

$$f(x) + u - L_1\{u\} \cong \sum_{j=1}^{N+L} \alpha_j \phi(r_j) + \alpha_{N+L+1} x + \alpha_{N+L+2} y + \alpha_{N+L+3}, \tag{12}$$

where $r_j = |x - x_j|$. The corresponding side conditions are given by

$$\sum_{j=1}^{N+L} \alpha_j = \sum_{j=1}^{N+L} \alpha_j x_j = \sum_{j=1}^{N+L} \alpha_j y_j = 0. \tag{13}$$

By forcing Eq. (12) to exactly satisfy Eqs. (7) and (13) at all nodes, we can get a set of simultaneous equations to uniquely determine the unknown coefficients $\alpha_j$. In this procedure, we need to evaluate the approximate particular solutions in terms of the RBF $\phi$. The standard approach is that $\phi$ in Eq. (11) is first selected, and then corresponding approximate particular solutions are determined by analytically integrating differential operator. The advantage of this method is that it is a mathematically reliable technique. However, this methodology easily performs only for simple operators and RBFs. Recently Muleskov et al. [5] made a substantial advance to discover the analytic approximate particular solutions for Helmholtz-type operators using the polyharmonic splines. But the analytical approximate particular solutions for general cases such as the MQ and other differential operators are not yet available now due to great difficulty involved. Another scheme evaluating approximate solution is a reverse approach [17,18]. Namely, the approximate particular solution is at first chosen, and then we can evaluate the corresponding $\phi$ by simply substituting the specified particular solution into certain operator of interest. It is a very difficult task to mathematically prove under what conditions this approach is reliable, although it seems to work well so far for many problems [17-19]. This scheme is in fact equivalent to the approximation of particular solution using the Kansa's method [8,9]. In this study, we use this scheme in terms of the MQ. The chosen approximate particular solution is

$$\varphi(r_j) = (r_j^2 + c_j^2)^{3/2}, \tag{14}$$

where $c_j$ is the shape parameter. The corresponding MQ-like radial function is

$$\phi(r_j) = 6(r_j^2 + c_j^2) + \frac{3r^2}{\sqrt{r_j^2 + c_j^2}} + (r_j^2 + c_j^2)^{3/2}, \tag{15}$$

Finally, we can get particular solutions at any point by weighted summation of approximate particular solutions at all nodes with coefficients $\alpha_j$. For more details on the procedure see [1-5].

## 3. Non-singular general solution and boundary knot method

One may think that the placement of source points outside domain in the MFS is to avoid the singularities of fundamental solutions. However, we found through numerical experiments that even if all source and response points were placed differently on physical boundary to circumvent the singularities, the MFS solutions were still degraded severely. In the MFS, the more distant the source points are located from physical boundary, the more accurate MFS solutions are obtained [2]. However, unfortunately the resulting equations can become extremely ill-conditioned which in some cases deteriorates the solution [2,6,7].

To illustrate the basic idea of the boundary collocation using nonsingular general solution, we take the 2D Helmholtz operator as an illustrative example, which is the simplest among various often-encountered operators having nonsingular general solution. The Laplace operator has not nonsingular general solution. For the other nonsingular general solutions see appendix.

The 2D homogeneous Helmholtz equation (8) has two general solutions, namely,

$$v(r) = c_1 J_0(r) + c_2 Y_0(r), \qquad (16)$$

where $J_0(r)$ and $Y_0(r)$ are the zero-order Bessel functions of the first and second kinds, respectively. In the standard BEM and MFS, the Hankel function

$$H(r) = J_0(r) + i Y_0(r) \qquad (17)$$

is applied as the fundamental solution. It is noted that $Y_0(r)$ encounters logarithm singularity, which causes the major difficulty in applying the BEM and MFS. Many special techniques have been developed to solve or circumvent this singular trouble.

The present BKM scheme discards the singular general solution $Y_0(r)$ and only use $J_0(r)$ as the radial function to collocate the boundary condition Eqs. (9) and (10). It is noted that $J_0(r)$ exactly satisfies the Helmholtz equation and we can therefore get a boundary-only collocation scheme. Unlike the MFS, all collocation knots are placed only on physical boundary and can be used as either source or response points.

Let $\{x_k\}_{k=1}^{N}$ denote a set of nodes on the physical boundary, the homogeneous solution $v(x)$ of Eq. (8) is approximated in a standard collocation fashion

$$v(x) = \sum_{k=1}^{N} \beta_k J_0(r_k), \qquad (18)$$

where $r_k = \|x - x_k\|$. $k$ is the index of source points. $N$ is the number of boundary knots. $\beta_k$ are the desired coefficients. Collocating Eqs. (9) and (10) in terms of Eq. (18), we have

$$\sum_{k=1}^{N} \beta_k J_0(r_{ik}) = b_1(x_i) - u_p(x_i), \tag{19}$$

$$\sum_{k=1}^{N} \beta_k \frac{\partial J_0(r_{jk})}{\partial n} = b_2(x_j) - \frac{\partial u_p(x_j)}{\partial n}, \tag{20}$$

where $i$ and $j$ indicate Dirichlet and Neumann boundary response knots, respectively. If internal nodes are used, we need to constitute another set of supplement equations

$$\sum_{k=1}^{N} \beta_k J_0(r_{lk}) = u_l - u_p(x_l), \qquad l = 1,\ldots,L, \tag{21}$$

where $l$ indicates the internal response knots and $L$ is the number of interior points. Now we get total $N+L$ simultaneous algebraic equations. It is stressed that the use of interior points is not always necessary in the BKM as in the DRBEM [15,17,20]. The term "boundary-only" is used here in the sense as in the DRBEM and MFS that only boundary knots are required, although internal knots can improve solution accuracy in some cases.

Before proceeding the numerical experiments, we consider to choose the radial basis function. In general, RBFs are globally defined basis functions and lead to a dense matrix, which becomes highly ill-conditioned if very smooth radial basis functions are used with large number of interpolation nodes [16]. This causes severe stability problems and computationally inefficiency for large size problem. A number of approaches have been proposed to remedy this problem such as domain decomposition and compactly supported RBFs (CS-RBFs). The latter is recently developed by Wendland [21], Wu [22] and Schaback [23]. Golberg et al. [2], Wong et al. [24] and Chen et al [25] respectively applied the CS-RBFs to the MFS, Kansa's method and DRBEM successfully. However, in this study we will not use the CS-RBFs to focus on the illustration of the basic idea of the BKM with globally-supported RBFs.

The MQ [26], TPS [27] and linear RBF [1] are most widely used globally-defined RBFs now. Among them, it is well known that the MQ ranks the best in accuracy [28] and is the only available RBF with desirable merit of spectral convergence [4]. However, its accuracy is greatly influenced by the shape parameter [29,30]. So far the optimal determination of shape parameter is still an open research topic. Despite this problem, the MQ is still most widely used in the RBF solution of various differential systems. For numerical example of the Laplace equation shown in the next section 4.2, the linear and generalized TPS RBFs show evidently slower convergence rate than the MQ. For example, the average relative error is 0.91% for the Linear RBF with 11 knots, 0.39% for the TPS with 11 knots, and 0.023% for the MQ (shape parameter 2) with 9 knots. We even got 0.5% relative average error by the MQ with only 3 knots. To simplify the presentation, this paper only uses the MQ with the DRM although the use of the TPS is also attractive in many cases.

## 4. Numerical results and discussions

In this paper, all numerical examples unless otherwise specified are taken from [20]. The geometry of test problem is all an ellipse featured with semi-major axis of length 2 and semi-minor axis of length 1. These examples are chosen since their analytical and numerical solutions are obtainable to compare. More complicated problems can be handled in the same BKM fashion without any extra difficulty. The zero order Bessel and modified Bessel functions of the first kind are evaluated via short subroutines given in [31]. The 2D Cartesian co-ordinates (x,y) system is used as in [20].

### 4.1. Helmholtz equation

The 2D homogeneous Helmholtz equation is given by

$$\nabla^2 u + u = 0 \tag{22}$$

with inhomogeneous boundary condition

$$u = \sin x. \tag{23}$$

It is obvious that Eq. (23) is also a particular solution of Eq. (22). Numerical results by the present BKM is displayed in Table 1 together with those by the DRBEM for comparison.

The numbers in the brackets of Table 1 mean the total nodes used. It is found that the present BKM converges very quickly. This demonstrates that the BKM enjoys the super-convergent property as in the other types of collocation methods [32]. The BKM solutions using 7 nodes are adequately accurate. In stark contrast, the DRBEM with 16 boundary and 17 interior points [20] produced relatively less accurate solution due to the use of the Laplacian and the low order of BEM convergence ratio. Please note that in this case there is no particular solution to be approximated by using the RBF and DRM in the BKM.

### 4.2. Laplace equation

Readers may argue that it is somehow unfair to choose the homogeneous Helmholtz equation to compare the BKM and DRBEM. The latter used the Laplace fundamental solution in the previous example. In the following we will further justify the superconvergence of the BKM through a comparison with the BEM for Laplace equation

$$\nabla^2 u = 0 \tag{24}$$

with boundary condition

$$u = x + y. \tag{25}$$

Eq. (25) is easily found to be a particular solution of Eq. (24). This homogeneous problem is typically well suited to be handled by the standard BEM technique. In contrast, there is inhomogeneous term in the BKM formulation to apply the nonsingular general solution of the Helmholtz operator. Namely, Eq. (24) is rewritten as

$$\nabla^2 u + u = u, \tag{26}$$

where the right inhomogeneous term $u$ is approximated by the DRM as shown in the section 2. The numerical results are displayed in Table 2 where the BEM solutions come from [20].

The MQ shape parameter $c$ is set 25 for both 3 and 5 boundary knots in the BKM. It is observed that the BKM solutions are not sensitive to the parameter $c$. It is seen from Table 2 that the BKM results using 3 boundary nodes achieve the accuracy of four significant digits and are far more accurate than the BEM solution using 16 boundary nodes. This striking accuracy of the BKM again validates its spectral convergence. In this case only boundary points are employed to approximate the particular solution by the DRM and RBF. It is noted that the coefficient matrices of the BEM and BKM are both fully populated. Unlike the BEM, however, the BKM yields symmetric coefficient matrix for all self-adjoint operators with one type of boundary conditions. This Laplace problem is a persuasive example to verify high accuracy and efficiency of the BKM vis-a-vis the BEM.

### 4.3. Convection-diffusion problems

The FEM and FDM encounter some difficulty to produce accurate solution to the systems involving the first order derivative of convection term. Special care need be taken to handle this problem with these methods. It is claimed that the BEM does not suffer similar accuracy problem. In particular, the DRBEM was said to be very suitable for this type problem [17,20]. Let us consider the convection diffusion equation

$$\nabla^2 u = -\partial u / \partial x, \tag{27}$$

which is given in [20] to test the DRBEM. The boundary condition is stated as

$$u = e^{-x}, \tag{28}$$

which also constitutes a particular solution of this problem. By adding $u$ on dual sides of Eq. (27), we

have

$$\nabla^2 u + u = u - \partial u/\partial x .\qquad(29)$$

The results by both the DRBEM and BKM are listed in Table 3.

The MQ shape parameter is chosen 4. The BKM employed 7 boundary knots and 8 or 11 internal knots. In contrast, the DRBEM [20] used 16 boundary and 17 inner nodes. It is stressed that unlike the previous examples, in this case the use of the interior points can improve the solution accuracy evidently. This is due to the fact that the governing equation has convection domain-dominant solution. Only by using boundary nodes, the present BKM with the Helmholtz non-singular solution and the DRBEM with the Laplacian [20] can not well capture convection effects of the system equation. It is found from Table 3 that both the BKM and DRBEM achieve the salient accurate solutions with inner nodes. The BKM outperforms the DRBEM in computational efficiency due to the super-convergent features of the MQ interpolation and global BKM collocation.

Further consider equation

$$\nabla^2 u = -\partial u/\partial x - \partial u/\partial y \qquad(30)$$

with boundary conditions

$$u = e^{-x} + e^{-y},\qquad(31)$$

which is also a particular solution of Eq. (30). The numerical results are summarized in Table 4.

In the BKM, the MQ shape parameter is taken 5.5. We employed 7 boundary knots and 8 or 11 inner points in the BKM compared with 16 boundary nodes and 17 inner points in the DRBEM [20]. The BKM worked equally well in this case as in the previous ones. It is seen from Table 4 that the BKM with fewer points produced almost the same accurate solutions as the DRBEM. Considering the extremely mathematical simplicity and easy-to-use advantages of the BKM, the method is superior to the DRBEM in this problem.

### 4.4. Varying-parameter Helmholtz problem

Consider the varying-parameter Helmholtz equation

$$\nabla^2 u - \frac{2}{x^2} u = 0 \tag{32}$$

with inhomogeneous boundary condition

$$u = -2/x. \tag{33}$$

Eq. (33) is also a particular solution of Eq. (32). This problem is the simplified Berger convection-diffusion problem given in [20]. Note that the origin of the Cartesian co-ordinates system is dislocated to the node (3,0) to circumvent singularity at $x=0$. The response knot-dependent nonsingular general solution of varying parameter equation (32) is given by

$$u(r_{ik}, x_i) = I_0\left(\frac{\sqrt{2}}{|x_i|} r_{ik}\right), \tag{34}$$

where $I_0$ is the zero order modified Bessel function of the first kind. $i$ and $k$ respectively index the response and source nodes. In terms of the BKM, the problem can be analogized by

$$\sum_{k=1}^{N} \alpha_k I_0\left(r_{ik}\sqrt{2}/x_i\right) = -2/x_i. \tag{35}$$

Note that only the boundary nodes are used in Eq. (35). After evaluating the coefficients $\alpha$, we can easily evaluate the value of $u$ at any inner node $p$ by

$$u_p = \sum_{k=1}^{N} \alpha_k I_0\left(r_{pk}\sqrt{2}/x_p\right). \tag{36}$$

Table 5 lists the BKM results against the DRBEM solutions. The BKM average relative errors under $N=9$, 13, 15 are respectively 9.7e-3, 8.1e-3 and 7.6e-3, which numerically demonstrates its convergence. The accuracies of the BKM and DRBEM solutions are comparable. It is noted that the DRBEM used 33 nodes (16 inner and 17 boundary knots) in this case, while the BKM only employed much less boundary knots. The accuracy and efficiency of this BKM scheme are very encouraging. Note that the present BKM representation differs from the previous ones in that we here use the response point-dependent nonsingular general functions. Similarly, we can easily constitute response node-dependent fundamental solutions. Thus, the essential idea behind this work may be extended to the BEM and DRBEM solution of varying parameter problems. For example, unlike the DRBEM scheme for varying velocity convection-diffusion problems given in Ref. [17], the variable convection-diffusion fundamental solutions with response node-dependent velocity parameters may be employed to the BEM or the DRBEM formulations, which may be especially attractive for high Peclet number problems.

## 5. Completeness concern

One of major potential concerns of the BKM is its completeness due to the fact that the BKM employs only the non-singular part of fundamental solutions of differential operators. This incompleteness may limit its utility. Although the given numerical experiments favor the method, now we can not theoretically ascertan of the general applicability of the BKM. On the other hand, Kamiya and Andoh [11] validated that the similar incompleteness occurs in the multiple reciprocity BEM using the Laplacian for Helmholtz operators. Namely, if the Laplace fundamental solution plus its higher-order terms are used in the MRM for the Helmholtz problems as in its usual form, we actually employ only the singular part of Helmholtz operator fundamental solution. Although the MRM performed well in many numerical experiments, it is mathematically incomplete. It is interesting to note that the BKM and MRM respectively employ the nonsingular and singular parts of the complete complex fundamental solution. It should also be stressed that although Power [13] simply indicated that the singular or nonsingular parts of Green representation can formulate the interior Helmholtz problems, no any related numerical and theoretical results are available from the published reports.

Kamiya and Andoh [11] also pointed out that the MRM formulation with the Laplacian can not satisfy the well-known Sommerfeld radiation conditions at infinity. Chen et al. [12] addressed the issues relating to spurious eigenvalues of the MRM with the Laplacian. Some literatures referred to in [12] also discussed the issues applying MRM with the Laplacian to problems with degenerate boundary conditions. These concerns of the MRM raise some cares concerning the applicability of the BKM which implements the nonsingular part of fundamental solution compared to the MRM using the singular part. Power [13] discussed the incompleteness issue of the MRM for the Brinkman equation and indicated that using one part of complex fundamental solution of Helmholtz operator may fail to the exterior Helmholtz problems. However, now we can not justify whether or not the BKM works for exterior problems since the method differs the MRM in using the DRM approximation of particular solutions. Dai [33] successfully applied the dual reciprocity BEM with the Laplacian to waves propagating problems in an infinite or semi-infinite region. It is worth pointing out that the Laplace fundamental solution used in the DRBEM also does not satisfy the Sommerfeld radiation condition. Unlike the MRM, the BKM and DRBEM do not employ the higher-order fundamental solutions to approximate the particular solution. Our next work will investigate if the BKM with the DRM can analyze the unbounded domain problems.

In fact, all existing numerical techniques encounter some limits. The BKM is not exceptional. Power [13] pointed out that the incompleteness in the MRM is problem-dependent. Therefore, the essential issue relating to the concerns of the BKM completeness is under what conditions the method works reliably and efficiently.

## 6. Concluding remarks

The present BKM can be regarded one kind of the Trefftz method [35] where the trial function is required to satisfy governing equation. The BKM distinguishes from the other Trefftz techniques such as the MFS in that we employ nonsingular general solution. The shortcomings of the MFS using fictitious boundary are eliminated in the BKM. The term "BKM" can be interpreted as a boundary modeling technique combining the DRM, RBF, and nonsingular general solution. In conclusion, the presented BKM is inherently possesses some desirable numerical merits which include meshless, boundary-only, integration-free, and mathematical simplicity. The implementation of the method is remarkably easy. The remaining two concerns of the BKM are the possible incompleteness in solving some types of problems due to the only use of nonsingular general solution and solvability of the Kansa's method for finding the particular solutions. More numerical experiments to test the BKM will be beneficial. This paper can be regarded a starting point of a series of works.

## Appendix

By its very basis, it is straightforward to extend the BKM to the nonlinear, three-dimensional, time-dependent partial differential systems. The following lists the nonsingular general solutions of some important steady and transient differential operators.

For the 3D Helmholtz-like operators

$$\nabla^2 u \pm \lambda^2 u = 0, \tag{A1}$$

we respectively have the nonsingular general solution

$$u^* = A\frac{\sin(\lambda r)}{r} \tag{A2}$$

and

$$u^* = A\frac{\sinh(\lambda r)}{r}, \tag{A3}$$

where sinh denotes the hyperbolic function, $A$ is constant. For the 2D biharmonic operator

$$\nabla^4 w - \lambda^2 w = 0, \tag{A4}$$

we have the nonsingular general solution

$$w^* = A_1 J_0(\lambda r) + A_2 I_0(\lambda r). \tag{A5}$$

The non-singular general solution of the 3D biharmonic operator is given by

$$w^* = A_1 \frac{\sin(\lambda r)}{r} + A_2 \frac{\sinh(\lambda r)}{r}. \tag{A6}$$

For the 3D time-dependent heat and diffusion equation

$$\Delta u = \frac{1}{k}\frac{\partial u}{\partial t}, \tag{A7}$$

we have the nonsingular general solution

$$u^*(r,t,t_k) = Ae^{-k(t-t_k)}\frac{\sin(r)}{r}. \tag{A8}$$

Furthermore, consider 3D transient wave equation

$$\Delta u = \frac{1}{c^2}\frac{\partial^2 u}{\partial t^2}, \tag{A9}$$

we get the general solution

$$u^*(r,t,t_k) = \left[ A_1 \cos(c(t-t_k)) + \frac{A_2}{c}\sin(c(t-t_k)) \right]\frac{\sin(r)}{r}. \tag{A10}$$

The 2D nonsingular general solutions of transient problems can be easily derived in a similar fashion. Two standard techniques handling time derivatives are the time-stepping integrators and the model analysis. The former involves some difficult issues relating to the stability and accuracy, while the latter is not very applicable for many cases such as shock. The BKM using time-dependent nonsingular solutions may circumvent these drawbacks. The difficulty implementing such BKM schemes may lie in how to satisfy the inharmonic initial conditions inside domain as in the time-dependent BEM. On the other hand, the time-dependent nonsingular general solutions may be directly applied in the domain-type RBF collocation schemes such as the Kansa's method. It is worth pointing out that the analogous method proposed by Katsikadelis et al. [15] may be combined with the BKM to handle the differential systems which do not include Laplace or biharmonic operators.

**Acknowledgements**:

Some valuable comments from a referee reshaped this paper into its present form. The authors also express grateful acknowledgement of helpful discussions with M. Golberg. Y.C. Hon and A.H.D. Cheng. The first author was supported as a JSPS fellow by the Japan Society of Promotion of Science.

**References**:

1. D. Nardini and C.A. Brebbia, A new approach to free vibration analysis using boundary elements. *Applied Mathematical Modeling*, **7** 157-162 (1983).

2. M.A. Golberg and C.S. Chen, The method of fundamental solutions for potential, Helmholtz and diffusion problems. In *Boundary Integral Methods - Numerical and Mathematical Aspects,* (Edited by M.A. Golberg), pp. 103-176, Computational Mechanics Publications, (1998).

3. C.S. Chen, The method of potential for nonlinear thermal explosion, *Commun Numer. Methods Engng*, **11** 675-681 (1995).

4. M.A. Golberg, C.S. Chen, H. Bowman and H. Power, Some comments on the use of radial basis functions in the dual reciprocity method, *Comput. Mech.* **21** 141-148 (1998).

5. A.S. Muleskov, M.A. Golberg and C.S. Chen, Particular solutions of Helmholtz-type operators using higher order polyharmonic splines, *Comput. Mech.* **23** 411-419 (1999).

6. T. Kitagawa, On the numerical stability of the method of fundamental solutions applied to the Dirichlet problem. *Japan Journal of Applied Mathematics*, **35** 507-518, (1988).

7. T. Kitagawa, Asymptotic stability of the fundamental solution method. *Journal of Computational*

*and Applied Mathematics*, **38** 263-269 (1991).

8. E.J. Kansa, Multiquadrics: A scattered data approximation scheme with applications to computational fluid-dynamics. *Comput. Math. Appl*. **19** 147-161 (1990).

9. E.J. Kansa and Y.C. Hon, Circumventing the ill-conditioning problem with multiquadric radial basis functions: applications to elliptic partial differential equations. *Comput. Math. Appls.* **39** 123-137 (2000).

10. Y.C. Hon and X.Z. Mao, A radial basis function method for solving options pricing model. *Financial Engineering*, **81**(1) 31-49 (1999).

11. N. Kamiya and E. Andoh, A note on multiple reciprocity integral formulation for the Helmholtz equation, *Commun. Numer. Methods Engng.*, **9** 9-13 (1993).

12. J.T. Chen, C.X. Huang, and K.H. Chen, Determination of spurious eigenvalues and multiplicities of true eigenvalues using the real-part dual BEM, *Comput. Mech.*, **24** 41-51 (1999).

13. H. Power, On the completeness of the multiple reciprocity series approximation, *Commun. Numer. Methods Engng.*, **11** 665-674 (1995).

14. W. Chen and M. Tanaka, New advances in dual reciprocity and boundary-only RBF methods, In *Proceeding of BEM technique conference,* (Edited by M. Tanaka), Vol. 10, pp. 17-22, Tokyo, Japan, (2000).

15. J.T. Katsikadelis and M.S. Nerantzaki, The boundary element method for nonlinear problems, *Engineering Analysis with Boundary Element*, **23** 365-273 (1999).

16. M. Zerroukat, H. Power and C.S. Chen, A numerical method for heat transfer problems using collocation and radial basis function, *Inter. J. Numer. Method Engng*. **42** 1263-1278 (1998).

17. L.C. Wrobel and D.B. DeFigueiredo, A dual reciprocity boundary element formulation for convection-diffusion problems with variable velocity fields, *Engng. Analysis with BEM*, **8**(6) 312-319 (1991).

18. N.A. Schclar, *Anisotropic Analysis Using Boundary Elements*, Comput. Mech. Publ., Southampton, 1994.

19. M. Kogl and L. Gaul, Dual reciprocity boundary element method for three-dimensional problems of dynamic piezoelectricity. In *Boundary Elements XXI*, 537-548, Southampton, 1999.

20. P.W. Partridge, C.A. Brebbia and L.W. Wrobel, *The Dual Reciprocity Boundary Element Method*, Comput. Mech. Publ., Southampton, UK (1992).

21. H. Wendland, Piecewise polynomial, positive definite and compactly supported radial function of minimal degree, *Adv. Comput. Math.* **4** 389-396 (1995).

22. Z. Wu, Multivariate compactly supported positive definite radial functions, *Adv. Comput. Math.* **4** 283-292 (1995).

23. R. Schaback, Creating surfaces from scattered data using radial basis function. In *Mathematical Methods for Curves and Surfaces*, (Edited by M. Dahlen et al.), pp. 477-496, Vanderbilt University Press, Nashville, (1995).

24. S.M. Wong, Y.C. Hon and M.A. Golberg, Compactly supported radial basis functions for the shallow water equations, *Appl. Math. Comput.* (to appear).

25. C.S. Chen, C.A. Brebbia and H. Power, Boundary element methods using compactly supported radial basis functions, *Commun. Numer. Meth. Engng.* **15** 137-150 (1999).

26. R.L. Hardy, Multiquadratic equations for topography and other irregular surfaces, *J. Geophys. Res.*, **176** 1905-1915 (1971).

27. J. Duchon, Interpolation des fonctions de deux variables suivant le principe de la flexion des plaques minces, *RAIRO Analyse Numeriques*, **10** 5-12 (1976).

28. R. Franke, Scattered data interpolation: tests of some methods, *Math. Comput.* **48** 181-200 (1982).

29. Y.C. Hon and X.Z. Mao, An efficient numerical scheme for Burgers' equation, *Appl. Math. Comput.* **95**(1) 37-50 (1998).

30. R.E. Carlson and T.A. Foley, The parameter $R^2$ in multiquadratic interpolation, *Comput. Math. Appl.* **21** 29-42 (1991).

31. W.H. Press, S.A. Teukolsky, W.T. Vetterling and B.P. Flannery, *Numerical Recipes in Fortran*, Cambridge University Press (1992).

32. C. Canuto, M.Y. Hussaini, A. Quarteroni and T.A. Zang, *Spectral Methods in Fluid Dynamics*, Springer, Berlin (1988).

33. D.N. Dai, An improved boundary element formulation for wave propagation problems, *Engng. Analysis with Boundary Element*, **10** 277-281 (1992).

34. J. Duchon, Splines minimizing rotation invariant semi-norms in Sobolov spaces, in: *Constructive Theory of Functions of Several Variables*, Springer-Verlag, Berlin, 1976.

35. R. Piltner, Recent development in the Trefftz method for finite element and boundary element application. *Advances in Engineering Software,* **2** 107-115 (1995).

Table 1. Results for Helmholtz problem

| x | y | Exact | DRBEM (33) | BKM (7) | BKM (11) |
|---|---|---|---|---|---|
| 1.5 | 0.0 | 0.997 | 0.994 | 0.999 | 0.997 |
| 1.2 | -0.35 | 0.932 | 0.928 | 0.931 | 0.932 |
| 0.6 | -0.45 | 0.565 | 0.562 | 0.557 | 0.565 |
| 0.0 | 0.0 | 0.0 | 0.0 | 0.0 | 0.0 |
| 0.9 | 0.0 | 0.783 | 0.780 | 0.779 | 0.783 |
| 0.3 | 0.0 | 0.296 | 0.294 | 0.289 | 0.296 |
| 0.0 | 0.0 | 0.0 | 0.0 | 0.0 | 0.0 |

Table 2. Results for Laplace problem

| x | y | Exact | BEM (16) | BKM (3) | BKM (5) |
|---|---|---|---|---|---|
| 1.5 | 0.0 | 1.500 | 1.507 | 1.500 | 1.500 |
| 1.2 | -0.35 | 0.850 | 0.857 | 0.850 | 0.850 |
| 0.6 | -0.45 | 0.150 | 0.154 | 0.150 | 0.150 |
| 0.0 | 0.0 | -0.450 | -0.451 | -0.450 | -0.450 |
| 0.9 | 0.0 | 0.900 | 0.913 | 0.900 | 0.900 |
| 0.3 | 0.0 | 0.300 | 0.304 | 0.300 | 0.300 |
| 0.0 | 0.0 | 0.0 | 0.0 | 0.0 | 0.0 |

Table 3. Results for $\nabla^2 u = -\partial u/\partial x$

| x | y | Exact | DRBEM (33) | BKM(15) | BKM (18) |
|---|---|---|---|---|---|
| 1.5 | 0.0 | 0.223 | 0.229 | 0.229 | 0.224 |
| 1.2 | -0.35 | 0.301 | 0.307 | 0301 | 0.305 |
| 0.0 | -0.45 | 1.000 | 1.003 | 1.010 | 1.000 |
| -0.6 | -0.45 | 1.822 | 1.819 | 1.822 | 1.818 |
| -1.5 | 0.0 | 4.482 | 4.489 | 4.484 | 4.477 |
| 0.3 | 0.0 | 0.741 | 0.745 | 0.744 | 0.743 |
| -0.3 | 0.0 | 1.350 | 1.348 | 1.353 | 1.354 |
| 0.0 | 0.0 | 1.000 | 1.002 | 1.003 | 1.004 |

Table 4. Results for $\nabla^2 u = -\partial u/\partial x - \partial u/\partial y$

| x | y | Exact | DRBEM (33) | BKM(15) | BKM (18) |
|---|---|---|---|---|---|
| 1.5 | 0.0 | 1.223 | 1.231 | 1.225 | 1.224 |
| 1.2 | -0.35 | 1.720 | 1.714 | 1.725 | 1.723 |
| 0.0 | -0.45 | 2.568 | 2.557 | 2.546 | 2.551 |
| -0.6 | -0.45 | 3.390 | 3.378 | 3.403 | 3.405 |
| -1.5 | 0.0 | 5.482 | 5.485 | 5.490 | 5.491 |
| 0.3 | 0.0 | 1.741 | 1.731 | 1.729 | 1.731 |
| -0.3 | 0.0 | 2.350 | 2.335 | 2.349 | 2.350 |
| 0.0 | 0.0 | 2.000 | 1.989 | 1.992 | 1.993 |

Table 5. Relative errors for varying parameter Helmholtz problem

| x | y | DRBEM(33) | BKM (9) | BKM (15) |
|---|---|---|---|---|
| 4.5 | 0.0 | 2.3e-3 | 3.3e-3 | 2.6e-3 |
| 4.2 | -0.35 | 2.1e-3 | 4.1e-3 | 3.3e-3 |
| 3.6 | -0.45 | 5.4e-3 | 6.8e-3 | 4.7e-3 |
| 3.0 | -0.45 | 4.5e-3 | 1.1e-2 | 4.4e-3 |
| 2.4 | -0.45 | 1.2e-3 | 1.4e-2 | 9.1e-4 |
| 1.8 | -0.35 | 9.0e-4 | 5.2e-3 | 1.7e-2 |
| 1.5 | 0.0 | * | 9.4e-3 | 3.4e-2 |
| 3.9 | 0.0 | 3.9e-3 | 7.0e-3 | 5.3e-3 |
| 3.3 | 0.0 | 3.3e-3 | 1.1e-2 | 6.3e-3 |
| 3.0 | 0.0 | 4.5e-3 | 1.3e-2 | 5.6e-3 |
| 2.7 | 0.0 | 2.7e-3 | 1.5e-2 | 3.4e-3 |
| 2.1 | 0.0 | 3.2e-3 | 1.6e-2 | 8.8e-3 |